\documentclass[aps,preprint,showpacs]{revtex4}
\font\rsfs=rsfs10.tfm scaled 1000
\usepackage{amsmath,amscd,amssymb,euscript,eufrak}
\usepackage{graphicx}
\usepackage{graphics}
\usepackage{epsfig}
\usepackage{ifthen}
\usepackage{verbatim}
\usepackage{color}

\def\bbf#1{\boldsymbol{#1}}

\def\c{_\mathrm{c}}

\DeclareMathAlphabet{\mathpzc}{OT1}{pzc}{m}{it}

\def\c #1{\mbox{\rsfs  #1}} %  kalig
\def\tsf#1{{\mathsf{ #1}}}

\def\vp{\varphi}

\def\Del{\Delta}          \def\del{\delta}

\def\ra{\rangle}          \def\la{\langle}

\def\beq{\begin{equation}}
\def\eeq{\end{equation}}
\def\beg#1#2{ \begin{#1}   #2 \end{#1}}
\def\beqm#1{ \beg{equation} {#1} }
\def\begg#1{ \begin{gather} #1 \end{gather} }

\def\SUM#1#2{\mbox{$\sum\limits_{#1}^{#2} $} }

\begin{document}
\def\ss #1{ \hskip -1.5pt #1 \hskip -1.5pt }
\def\pl{\partial}
\def\ssquare{{\scriptscriptstyle \square}}
\def\ssq{{\scriptscriptstyle \square}}
\title{
Thomson rings in a disk
}
\author{M. Cerkaski$^{1}$, R.G. Nazmitdinov$^{2,3}$, and A. Puente$^{2}$}
\affiliation{$^{1}$Institute of Nuclear Physics PAN, Department of Theory
of Structure of Matter, 31-342 Cracow, Poland\\
$^{2}$ Departament de F{\'\i}sica,
 Universitat de les Illes Balears, E-07122 Palma de Mallorca, Spain\\
$^{3}$Bogoliubov Laboratory of Theoretical Physics,
Joint Institute for Nuclear Research, 141980 Dubna, Russia
}
\begin{abstract}
We discuss the basic principles of self-organization of a finite number of
charged particles interacting via the
1/r Coulomb potential in disk geometry.
The analysis is based on the cyclic symmetry and
periodicity of the Coulomb interaction between particles located on several rings.
As a result, a system of equations is derived, which allows us readily to determine with
high accuracy the equilibrium configurations
of a few hundreds charged particles.
For $n\gtrsim 200$ we predict the formation of a hexagonal core and
valence circular rings for the centered configurations.
\end{abstract}
\date{\today}
\maketitle
%05.65.+b 	Self-organized systems (see also 45.70.-n in classical mechanics of discrete systems)
%45.50.Jf 	Few- and many-body systems
%41.20.Cv 	Electrostatics; Poisson and Laplace equations, boundary-value problems
%82.70.Dd 	Colloids
%64.75.Yz 	Self-assembly
%36.40.Wa 	Charged clusters
The distribution of charged particles
on a two-dimensional curved surface,
considered first by Thomson \cite{tom},
has attracted continuous attention for
a decade \cite{1}.
This problem provides useful insights
into the physics of quantum dots and Bose-Einstein condensates
\cite{fin}, topological defects \cite{koul,mug,yao}, and
colloidal systems, where
colloidal particles self-assemble at the interface
of two distinct liquids  such as
particle--stabilized \cite{bin} or
charged--stabilized emulsions \cite{col1,col2}.

Considering the electron distribution in
a circular harmonic oscillator classically, Thomson
found that interacting
electrons are self-assembled in a family of rings (shells)
with a specific number of electrons
due to equilibrium conditions.
Thirty years later, Wigner \cite{wig} predicted the formation
of an electron lattice in an infinitely three-dimensional (3D) extended system at low density.
These problems have common roots related to the dominance of the Coulomb interaction
over the kinetic energy.
Both models play a major role in our understanding of equilibrium configurations of
interacting  particles in the case of a soft confinement and in the absence of confinement.
Evidently, however, they are different  with respect to the role played by
the number of particles, boundary conditions and symmetry.
For an infinitely large box  the discrete translation symmetry is responsible for
the ordered structure in the Wigner crystal.
In a circular trap, with a finite number of electrons,
the cyclic symmetry gives rise to the formation of shells.
For finite systems the
role of confinement and its underlying symmetries are crucial for the formation of
equilibrium configurations \cite{bir}.

Thanks to modern technology many ideas and concepts developed early
can be analysed with high accuracy.  Recent experimental studies
of the additional electron energies of small number of electrons in a trap over
a liquid-helium film \cite{rous} confirm the results obtained by means of
classical Monte-Carlo calculations for the harmonic oscillator
trap \cite{pet1,loz,bol}. The results demonstrate
that $n$ point charges located on a ring
create equidistant nodes  as predicted by Thomson \cite{tom}.
There are hundreds of papers on the  self-organization of charged particles in disk geometry
(a hard confinement) in different fields of physics and chemistry (see, for, example, 
\cite{fin,yao}) where various  simulation techniques are used.
Although a similar pattern is obtained for a hard wall potential for
$n\leq 50$ (c.f., \cite{pet2}),
the distribution of particles
is very different from the one found for the harmonic oscillator confinement.
Such a deviation, noticed already a few decades ago \cite{ber}, is not  understood yet.
Indeed, the results of numerical simulations are rather formal, because they are not
based on any well established model,
while neither the Thomson nor the Wigner model mentioned above is relevant there.
In contrast to the harmonic oscillator case, a consistent analysis
of the shell pattern obtained by simulation techniques in disk geometry has been
lacking up to now (for a review, see \cite{wor}). Among the latest developments,
we could mention
the approach  based on a continuum limit \cite{koul,mug}.
Although this approach  describes a general trend of the density distribution in
the framework of elasticity theory, it is unable
to provide a detailed description of the shell structure for a finite number of particles.

In this paper, we present a model that enables one to
describe with high accuracy the ground state configuration
of charged particles in a disk
as a function of particle number.  Although we consider the classical system at zero temperature,
our approach could shed
light on the nature of self-organization of colloidal particles in organic solvents,
charged nanoparticles absorbed at oil-water interfaces, and electrons trapped on
the surface of liquid helium.
To address the problem, we consider particles (electrons)
confined in a planar disk and interacting via the Coulomb interaction.
To check the validity of our theoretical approach,
we also perform molecular dynamics (MD)
calculations similar to the one discussed in \cite{wor} and compare
our predictions with the MD
results for $n\leq 400$ particles.

The MD results indicate  that for $n\leq 11$ the
equilibrium configuration is defined
by all particles equally distributed on the circle with radius $R$.
In this case the minimal energy of the system is
\begin{eqnarray}
\label{Ering}
&&E_n(R)=\frac{\alpha}{2\,R}\sum_{i=1}^{n-1}\sum_{j=i+1}^{n}\frac{1}{\sin\frac{\pi}{n}(|i-j|)}=
\frac{\alpha nS_n}{4\,R}\,,\\
\label{sn}
&&S_n=\sum_{k=1}^{n-1}\frac{1}{\sin \frac{\pi}{n}k}\,.
\end{eqnarray}
Here, $\alpha=e^2/4 \pi \varepsilon_0 \varepsilon_r$. Below, for the sake of discussion, we
use $\alpha=1$, unless stated otherwise.
We recall, that with the harmonic oscillator confinement,
already for $n=6$ one obtains one particle at the center  $(5+1)$\cite{tom,pet1,loz,bol}.

Let us suppose that the system is stable with $n$ particles located
at the circle boundary. If we add a particle then:
either i) it is placed at the
boundary with a total energy $E_{n+1}$; or ii) due to circular symmetry,  it is located
at the center, interacting with the external $n$ charges, and the total energy is $E_{n}(R) + n/R$.
The critical number of charged particles for this transition is defined by the condition
$(E_{n} (R) + n/R)-E_{n+1} (R) \le 0$,
 which yields the following equation
\begin{equation}
(n+1)S_{n+1}\geq nS_n+4n\,.
\end{equation}
The resolution of this equations provides the critical number $n=11$.
In other words,
eleven charged particles lie on the circle boundary, while for twelve charged particles
there are eleven charged particles at the boundary, and one is located at the
centre.

For $n\geq 12$ the MD calculations show the formation of several internal rings.
In particular, for $n\leq 29$ the number of electrons grows in two rings
until two complete shells $(23+6)$ are formed.
Evidently, the interaction between electrons from different rings should be included now.
To obtain further insight into the formation of the equilibrium configuration,
we consider the Coulomb interaction between two
rings with radiuses $r_1$, $r_2$, and  $n$ and $m$ electrons, respectively, uniformly distributed on each ring.
Thus, we have
\begg{  \label{en}  E_{nm}(r_1,r_2,\psi) \ss= \SUM{i=1}n\,\SUM{j=1}m\, \epsilon (r_1,r_2, \psi^{nm}_{ij}+\psi)\,,\\
\label{eijnm} \epsilon (r_1,r_2,\theta)= (r_1^{2}+r_2{^2}-2\,r_1\,r_2\,\cos \theta)^{-1/2}\,, }
where $ \psi^{nm}_{ij}= 2 \,\pi(i/n -j/m)$ and $\psi$ is the relative angular offset between the two rings.

It can be shown that the set of $n \times m$ angles  $ \psi^{nm}_{ij}$ is equivalent, in the interval $[0, 2\pi]$, to the $G$-fold set
$\{\psi_k= 2\pi/L\times k, k=1,\ldots, L\}$. Here $L\equiv {\rm LCM}(n,m)$ and
$G\equiv{\rm GCD}(n,m)=n\times m/L$ are the  least common multiple and
greatest common divisor of the  of numbers $(n,m)$, respectively.
As a result  Eq.(\ref{en})  transforms to
\beqm{
 \label{en1}
 E_{nm}(r_1,r_2,\psi) ={G} \SUM{k=1}{L} \epsilon (r_1,r_2,\psi_k +\psi)\,,
}
which can be applied to any $ 2\pi$ periodic function $\epsilon (r_1,r_2,\theta)$.
In turn, this result shows that these kind of functions
are invariant under angle transformations corresponding
to the cyclic group of $L$ elements, implying a $\Delta_{nm}=2\pi / L$ periodicity
 \beqm{
 E_{nm}(r_1,r_2,\psi+\Delta_{nm})=E_{nm}(r_1,r_2,\psi).}
This is a key result of our approach which allows us  to simplify drastically
the problem of equilibrium configurations and underlines the importance of the cyclic symmetry.

By virtue of the fact that
the ring-ring interaction is an even periodic function in the angle $\psi$, it can be presented
by means of a Fourier series of cosines,
\begg{
\label{fser}
E_{nm}(r_1,r_2,\psi)=\la E_{nm} \ra \! +\! \SUM{\ell=1}{\infty}C_{\ell nm}(r_1,r_2)\cos(\ell L \psi).}
The average value is obtained by integrating in $\psi$, and 
using Eq.(\ref{en1}) we have
\begg{
\la  E_{nm} \ra=\frac{1}{2\pi}\int_0^{2\pi}\!\!\!\!\! d\psi \, E_{nm}(r_1,r_2,\psi) \nonumber\\ 
=\frac{G}{2\pi}\SUM{k=1}{L}\int_{0}^{2\pi} \!\!\!\!\! d\psi \,
\epsilon(r_1,r_2,\psi_k+\psi)\, .
\label{aven}}
All terms in the sum Eq.(\ref{aven}) give the same contribution, and we obtain in terms of the complete elliptic integral of first kind  \cite{AS}
\beqm{
\la  E_{nm} \ra=\frac{2nm}{\pi r_>(1+t)} K(4 t/(1+t)^2) =2nm\frac{K(t^2)}{\pi r_>}\, .
\label{aven2}}
Here, we introduced notations: $r_>=\max(r_1,r_2)$, $r_<=\min(r_1,r_2)$, $t=r_</r_>$;
and  used the symmetry property
$K(4t/(1+t)^2)  = (1 + t)\,  K(t^2)$.
It is noteworthy that the average value  $\la E_{nm} \ra$ is exactly the interaction energy
between homogeneously distributed $n$ and $m$
charges over first and second rings, respectively.

In a similar way, the Fourier coefficients are given by
\begg{
C_{\ell nm}(r_1,r_2)=\frac{1}{\pi}\int_0^{2\pi} \!\!\!\!\! d\psi \, \cos(\ell L \psi) E_{nm}(r_1,r_2,\psi)\nonumber\\
=\frac{nm}{\pi}
\int_0^{2\pi} \!\!\!\!\! d\psi \, \frac{\cos(\ell L\psi)}{\left[r_1^2+r_2^2-2r_1r_2\cos \psi \right]^{1/2}}\,.
\label{clm}
}
By means of series expansion, it can be shown that
\beqm{
\label{cm12}
C_{\ell nm}(t)\approx\frac{nm}{r_>}d_{\ell L}t^{\ell  L}
+O(t^{\ell L+2})\,,
 }
where $d_{M}\!=\! 2 (2M-1)!! / (M! \, 2^M)$ is a slowly decreasing coefficient. Evidently,
at large $L$ the contribution brought about by
the fluctuations related to the $\psi$ variable in the series (\ref{fser}) is very small,
even for the first harmonic $\ell=1$.
For illustration, we consider the fluctuating part of the ring-ring energy
\begg{ \label{del} \Del E_{nm}(R,r) =   \la E_{nm}\ra -  E_{nm}(R,r,\psi=\pi/L)
}
for
$m=6$ in the internal ring with a radius $r$ and
vary the electron number $20\leq n\leq 25$  in the external ring with a radius $R$.

\begin{figure}[ht]
\vspace{-0.5cm}
\includegraphics[scale=.45]{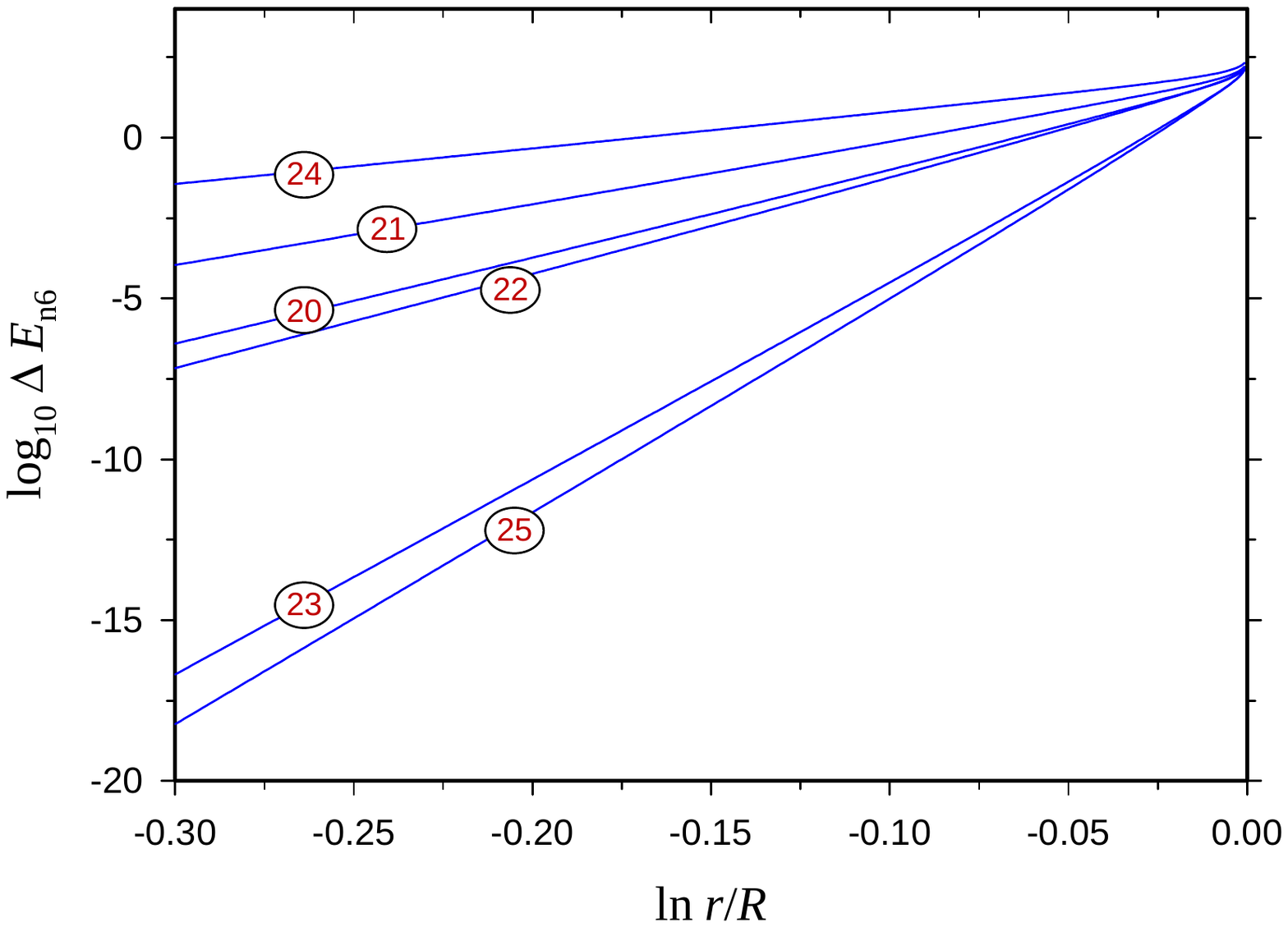}
\vspace{-7.6cm}
  \caption{ The fluctuation of ring-ring energy, Eq.(\ref{del}),  as a function
  of the ratio $x=r/R$ for $m=6$ in the internal ring and
  $20\leq n\leq 25$ in the external ring.}
   \label{DISCdum}
 \end{figure}

From the numerical analysis (see Fig.\ref{DISCdum}) it follows that
\beqm{\label{est}  \Del E_{nm}(R,r)   \sim   c_L\,(r/R)^L,    }
which is fully  consistent with the estimation (\ref{cm12}).
It is evident that fluctuations
hardly play a role in the ring-ring interaction at large $L$, for instance,
when $n\!=\!23 \,(L\!=\!138)$, or $n\!=\!25 \,(L\!=\!150)$.  Even
for the worst case $n\!=\!24 \,(L\!=\!24)$, it amounts to a small fraction of  $\la E_{nm}\ra$ for ratios $r/R < 0.8$.

The results for two interacting rings guide us to tackle the issue of the total energy.
The total energy of $n$ charged particles in a disk of radius $R$ is
\beqm{ \label{Etot}  \c E_{\rm tot} (\tsf{n},\tsf{r},\bbf{\vp})\ss=
       \SUM{i\ss=1}{p} E_{n_i}(r_i) + \! \SUM{i<j}{p} E_{n_i n_j}(r_i,r_j,\vp_i-\vp_j).}
Here, $\tsf{n}\!=\!(n_1,\ldots,n_p)$ is a partition of the total number $n$ on
$p$ rings with radiuses $\tsf{r}\!=\!(r_1,\ldots,r_p)$ and offset angles  $\bbf{\vp}\!=\!(\vp_1\!=\!0,\ldots,\vp_p)$.
We assume $R=r_1=1 > r_2 > \cdots > r_p$.  The results for two rings suggests to define the
total energy as $\c E_{\rm tot}(\tsf{n},\tsf{r},\bbf{\vp})=\c E_{\rm avg}(\tsf{n},\tsf{r})+\delta E(\tsf{n},\tsf{r},\bbf{\vp})$ with
\beqm{\label{Ecoll} \c E_{\rm avg} (\tsf{n},\tsf{r}) =  \SUM{i=1}{p}\,n_i\frac{S_{n_i}}{4r_i}+
\frac{2}{\pi} \SUM{i<j}{p} \,n_i\,n_j\,\frac{K ( (r_j/r_i)^2) }{r_i}\,,}
and neglect the dependence on
the relative angles $\psi=\vp_i-\vp_j$ , i.e., the term $\delta E(\tsf{n},\tsf{r},\bbf{\vp})$.

The equilibrium configuration of particles can be obtained by minimizing the energy (see Eq.(\ref{Ecoll}))
with respect to $(p,\tsf{n},\tsf{r})$,
i.e., finding the partition corresponding to the lowest total energy. For a given partition, the set of equations
$r_i\partial \c E_{\rm avg} (\tsf{n},\tsf{r})/\partial r_i=0$
that determines the optimal radiuses $(r_i, i=2,\ldots,p)$ is
\begg{    \label{req}  \,r_i^{2} \SUM{j=i+1}{p}{}\, \frac{n_j\,{\rm E}( (r_j/r_i)^2)}{r_j{^2} - r_i^2}\\
  +     \,r_i\,\SUM{j=1}{i-1}{} n_j\,\Bigl( \frac{ r_j\,{\rm E}((r_i/r_j)^2)}{r_j^2 - r_i^2} -
   \frac{ {\rm  K}((r_i/r_j)^2)}{r_j}\Bigr)     = \frac{\pi}{8}S_{n_i}. \nonumber}
 Here, ${\rm E}$ is the complete elliptic integral of the second kind. 
 Thus, instead of searching for the optimal arrangement of $n$ particles by means of simulation techniques,
 one must seek the partition $\tsf{n}$ which provides the lowest value of $\c E_{\rm avg}$ by solving a system of a few ($p-1$) equations.

Numerical analysis of the system (\ref{Ecoll}),\!\! (\ref{req}), shows that, once one electron appears at the center,
it gives rise to a new internal ring (shell) which is progressively filling with electrons.
The list of lowest energy configurations with filled shells reads,
$n/\c E_{\rm avg} \{ \tsf{n} \}: 11/48.5757 \{11\} ; 29/444.548 \{23, 6\}; 55/1792.01 \{37, 13, 5\}; $
$90/5115.56 \{53, 20, 12, 5 \}; 135/11995.4 \{70,29,19,12,5 \}$.
The largest number of electrons lies on the circle boundary and
 decreases with sequential access to inner shells $2,3,\ldots$.
The numerical solution of the system (\ref{Ecoll}),\!\! (\ref{req}), provides
a remarkable agreement
with the MD calculations for equilibrium configurations up to $n=105$,
excluding a few cases (see Table \ref{tb1}). Our MD results agree with those of Ref.\,\onlinecite{wor} up to 
$n\!=\!160$ particles, while we obtain lower energies for $n\!=\!400,500,1000$ and also systematically better 
values for $n>52$ than those implied in Fig. 8 of Ref.\,\onlinecite{mug}.
\begin{table}[htb]
\caption{Values for the only seven cases where optimal configurations, obtained with the aid of Eq. (\ref{req}), disagree with
the MD results. The MD results can be found also in \cite{wor}.
}
\begin{tabular}{|c|c|c|c|c|}
\hline
$n$  & $\c E_{\rm avg}(n)$ & $\del$ &  {\rm configuration} \\
\hline
 38  &  805.021         &  -0.101404  & $(28, 9, 1)_2^3$   \\
  61 &  2237.25         &  -0.056784  & $(39, 14, 7, 1)_3^1$  \\
  76 &  3575.38         &  -0.176466  & $(46, 17, 10, 3)_3^1$ \\
  79 &  3881.59         &  -0.164677  & $(48, 17, 10, 4)_4^2$ \\
  88 &  4878.17         &  -0.109206  & $(51, 20, 12, 5)_3^1$ \\
  90 &  5115.56         &  -0.155515 & $(53, 20, 12, 5)_1^5$ \\
  97 & 5991.62          &  -0.148982  & $(55,21, 13, 7, 1)_4^2$ \\
\hline
\end{tabular}
\label{tb1}
\end{table}

The difference $\delta= E_{\rm MD}-\c E_{\rm avg}$  provides the error of our approximation.
The rings are counted starting from the external one
which is the first ring. The notation $(28, 9, 1)_2^3$  means that we have to add
one particle in the third ring and remove one particle from the second ring in order
to obtain the MD result. Although the total energy errors are very small, the assumptions of our model fail to predict the
correct configurations for the shown total $n$. The reason for this is twofold.  First,
as discussed above,  the fluctuating part (see Eqs.(\ref{fser}), \!\!(\ref{clm}))
diminishes when $L$ is large, while at small $L$ its
contribution may affect the prediction of the optimal configuration. Second, some MD configurations slightly break circular symmetry,
which is not considered in the present approach. 
Nevertheless, we stress that in practical cases considered so far for $n \leq 400$, the solution of our equations (\ref{Ecoll}),\!\! (\ref{req})
reduces the CPU time by a considerable factor (about $10^3$ for $n\approx 400$) in comparison with the MD calculations. 
Moreover, with the aid of this solution as a guide for the initial MD particle positions
one reduces drastically the scanning effort to find the exact ground state configurations. 
We recall that systematic studies of equilibrium configurations in a disk geometry 
with Monte Carlo simulation techniques and MD calculations have been done up to 
$n\leq 50$ \cite{pet2} and $n\leq 160$ \cite{wor}, respectively.

Starting from $n=106$,
the predictions based on the energy $\c E_{\rm avg}$ and MD results demonstrate
a systematic  $\Delta n\approx |2|$ disagreement in the partition of charged particles between
available rings.  In particular, the particle number, which corresponds to
the opening of a new shell (starting from one particle in the center),
can be  calculated exactly up to $n=90$ with the aid  of the formula
$n=(2p+1)(2p+2)$ (see also \cite{tur}).
It gives $n=132$ at $p=5$, while the MD results
provide the opening of the sixth shell at $n=134$. Our calculations predict
this opening at $n=136$.  Nevertheless, by means of this formula one obtains
an estimation of the shell number $p$ associated with a given particle number $n>90$:
$p \simeq \sqrt{n}/2$.

The increase in the particle number gives rise to the onset
of a centered hexagonal  lattice (CHL) at the core of the disk 
(see also discussion in \cite{pet2,wor,mug}).
 In fact, for equilibrium configurations close to the one which opens
a new shell, we find an increasing sequence of rings, starting from the center, with $n_k = 6 \, k$ particles
matching the regular hexagonal pattern.
This is clearly seen in the results for  $n\!=\! 92 \,  \{ {\bf 1}, {\bf 6}, {\bf 12}, 20,53 \}$,
$n\! =\! 136  \, \{ {\bf 1}, {\bf 6}, {\bf 12}, 19, 28, 70\}$,  $n\! = \!187  \,  \{ {\bf 1}, {\bf 6}, {\bf 12}, {\bf 18}, 26, 37, 87\}$, $\ldots$,
$n \! = \! 395  \,  \{ {\bf 1}, {\bf 6}, {\bf 12}, {\bf 18}, {\bf 24}, 32, 40, 50, 65, 147\}$. It is worth mentioning that the relative error in
$\c E_{\rm avg}=110667.6$ for $n=395$ with respect to the MD result (=110665.1) is only $2 \, \times 10^{-3} \%$.
For even bigger systems we find the formation  of just seven full shells, $n \! = \! 1976  \,  \{ {\bf 1}, \ldots , {\bf 42}, \ldots\}$ before the
symmetry of the circular confining geometry takes over.

\begin{figure}[ht]
\vspace{-1.6cm}
\hspace{-0.75cm}
\includegraphics[scale=.234]{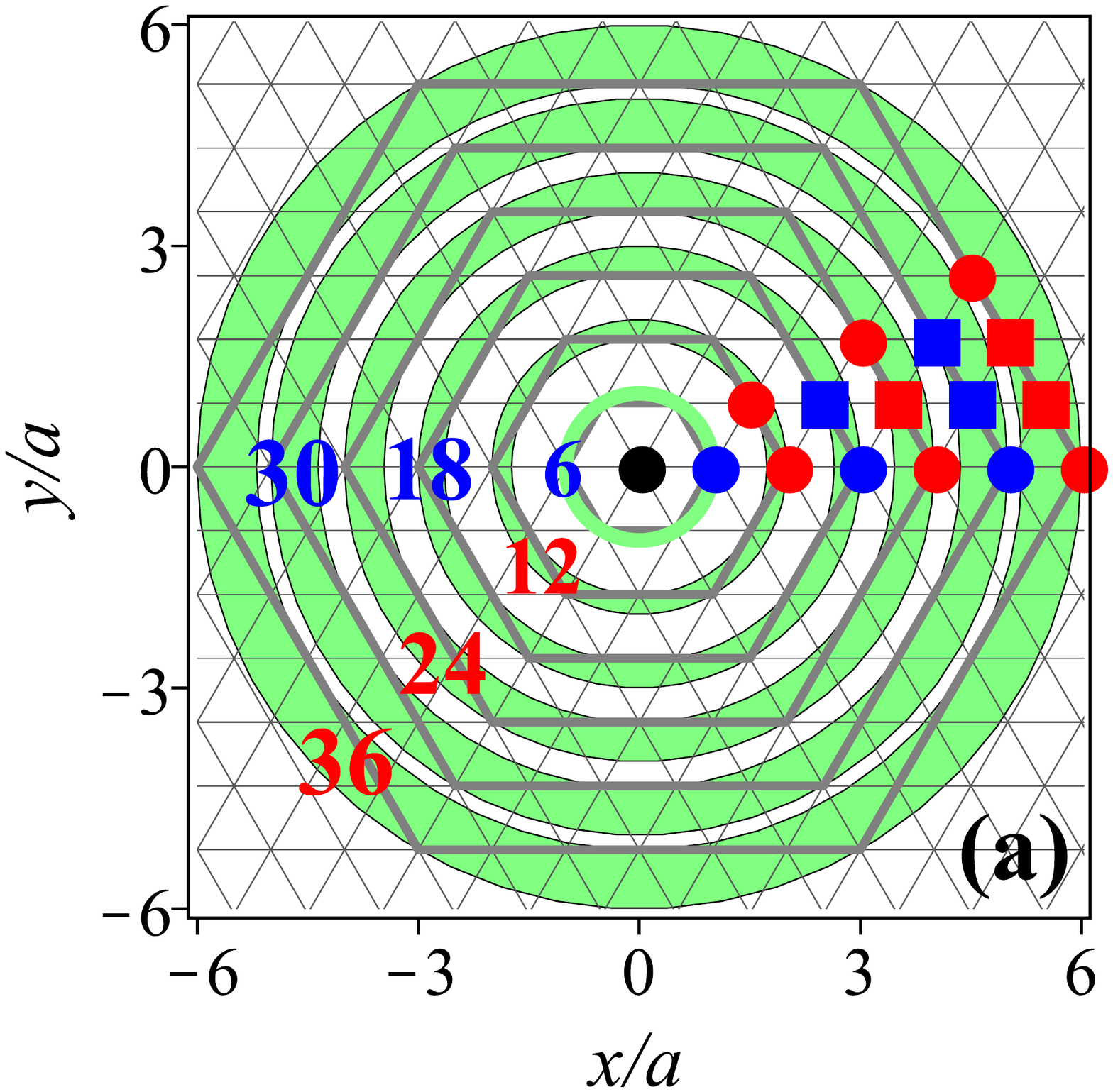}
\hspace{-0.71cm}
\includegraphics[scale=.231]{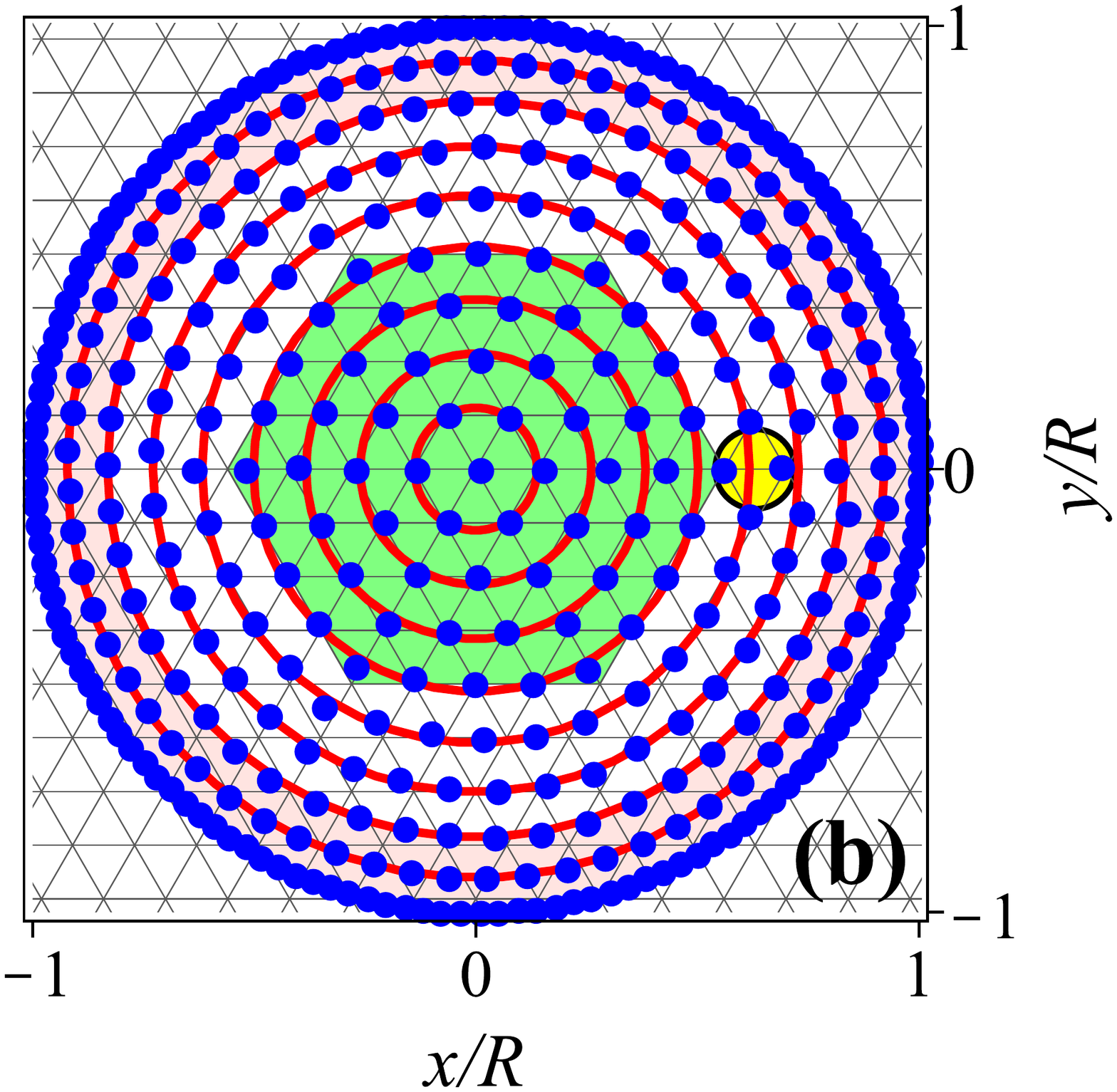}
\vspace{-1.5cm}
 \caption{(Color online)
  (a): Structure of internal ({\it core}) rings corresponding to the CHL. Each shell (green) contains a family of circles
  with radii $R_{kl}$ and particle numbers $n_{\rm ring}=6\, p$ (see text).
  Solid dots and squares correspond to lattice sites with 6- and 12-fold multiplicity, respectively.
 (b): Comparison of the numerical solution of Eq.(\ref{req}) (rings) with
 the MD results (dots) for $n=395$ particles. The {\it core} (green) region with
 ${\{\bf 1}, {\bf 6}, {\bf 12}, {\bf 18}, {\bf 24}\}$ particles, exhibits a hexagonal pattern.
 The five external  {\it valence} shells contain
  $147, 65, 50, 40, 32$ particles with an almost perfect circular structure for the three outer rings (pink). There is a
  small mismatch, involving two particles at the intermediate region, displayed within the small (yellow) circle.}
   \label{ss}
 \end{figure}

This fact can be understood by considering the arrangement of the
CHL points,  $\vec{x}_{k,\ell}=k\vec{a}_1+\ell \vec{a}_2$, given by integers $k, \ell$ and the two primitive Bravais lattice vectors $\vec{a}_1=a(1,0)$ and
$\vec{a}_2=a(1/2,\sqrt{3}/2)$, where $a$ is the lattice constant. The $n_p=6 \, p$ sites in the $p-{\rm th}$ shell are organised
in different circular rings with radii $R_{k \ell}\ss=a\,\sqrt{k^2+ \ell^2+k\,\ell}$, where $p=k + \ell$ and $0\leq \ell \leq k$,
containing either 6 (if $\ell=0, k$) or 12 (otherwise) particles (see Fig.\ref{ss}a). Up to $p=7$ all these radii are well ordered within and between
successive shells, and the model we presented groups them in a single circular shell $n_{\rm ring}=6\, p$. Beyond the
seventh shell, however, rings start to overlap (e.g. $R_{7,0} > R_{4,4}$), ultimately distorting this sequence as
they depart from the center.

\begin{figure}[ht]
\includegraphics[width=0.82\textwidth]{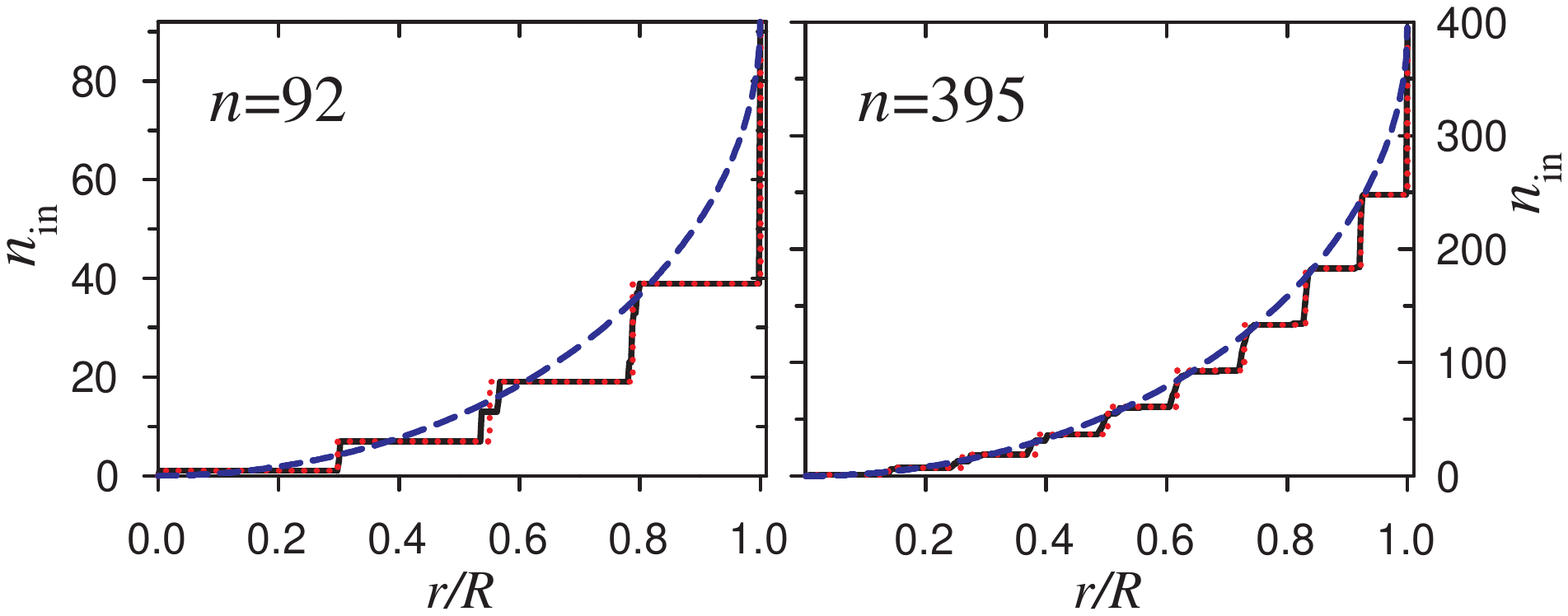}
\vspace{-14cm}
%\vspace{-1.5cm}
 \caption{(Color online) Number of charged particles within a disk of radius $r$.
 The results of MD, our model and a continuum model \cite{mug} are
 shown respectively by a solid (black), dotted (red), and dashed (blue) line for $n\!=\!92$ and $n\!=\!395$ charges.}
  \label{sp}
 \end{figure}

A comparison of the predictions of our model with the MD results for $n=395$ particles is shown in  Fig.\ref{ss}b.
The discrete equilibrium positions at the core of the structure are nicely located over a hexagonal lattice that gets progressively distorted as one moves towards the boundary, where particles are arranged in almost perfect circular shells. As discussed above, we consider
the interaction of homogeneously distributed charges on several rings,
neglecting the angular displacement between them. This first order approximation hides the mechanisms of
topological defects (see Fig.\,\ref{ss}b, small (yellow) circle) discussed, for example,
by Mughal and Moore \cite{mug} by means of a continuum model.
This model neglects, however, finite size fluctuations and is only appropriate for very large systems. 
In contrast, our model is able to reproduce the shell pattern
obtained with MD calculations for any finite $n$ remarkably well (see Table\,1, Figs.\,\ref{ss}b, \ref{sp}). 
It is interesting to note that the number of charges in the outer shell (ring) fitted 
to our model data ($n\leq 400$)
is well reproduced by the formula $n_{\it out}=2.795\, n^{2/3}-3.184$ and 
confirms the power law scaling obtained also in Refs.\onlinecite{wor,mug}.
%and the continuum model \cite{mug}.  
Similar scalings can also be found for subsequent shells as well as for the smooth part
of the total energy. However, these fits should be taken cautiously, since the coefficients depend 
strongly on the fitting range and the quality of the data, which is assumed to correspond 
to the lowest energy configuration. Further 
refinement of our method brought about by including the angular dependence and understanding of phenomenological coefficients requires a dedicated study
 and is a subject of a forthcoming paper.

Increasing  the particle number at fixed $R$, one reaches the situation
in which quantum corrections due to electron zero-point fluctuations, $(\Delta r)^2$, around the equilibrium position become important. 
To quantify this effect we use the de Boer parameter $\Lambda=(\Delta r)^2/a^2$ \cite{hom}, with $a=R/p\simeq 2R/\sqrt{n}$ being the mean inter-particle separation, far from the boundary. 
As a rough estimate for $(\Delta r)^2$ we consider the harmonic approximation to the potential seen
by a particle at the center
($r_p=0$) of the structure, which can be expanded as
\begin{eqnarray}
&&V(r)=2\alpha/\pi\times\sum_{i=1}^{p-1} n_i K((r/r_i)^2) / r_i\nonumber\\
&&=\alpha/R\times(V_0+\frac{1}{4} V_2 \, (r/R)^2+\cdots),
\end{eqnarray}
 with
$V_k =\sum_{i=1}^{p-1} {n_i}/{(r_i/R)^{k+1}}$. The coefficients $V_k$ are size dependent and may be generally fitted by a series in $\sqrt{n}$.
In particular, considering all equilibrium configurations with one particle at the center for $n \leq 400$ we obtain
$V_2 \simeq A_2 \,n^{3/2} \, (1+O(1/\sqrt{n})$ with $A_2 \approx 0.625$. In the harmonic approximation one then has
\begin{equation}
m \omega^2 R^2=\alpha V_2 /(2R)\,, (\Delta r)^2=\hbar / m\omega,
\end{equation} 
providing the following estimate
\begin{equation}
\Lambda^2=\hbar^2/(m\alpha) \times \sqrt{\pi \sigma}/(8 A_2)
\end{equation} 
in terms of the particle density $\sigma=n/\pi R^2$. Quantum melting is avoided
at $\Lambda \leq \Lambda_0  \sim 0.2$ \cite{hom}, corresponding to an upper bound for the density, $\sigma_0$.
As an example we obtain for electrons in GaAs ($m=0.067 \,m_e, \varepsilon_r \approx 12.4$) and $R=1 \mu m$
 the onset of  {\it cold melting} at $n \gtrsim 410$ particles, for which quantum corrections should be 
necessary.

In conclusion, we have derived a system of equations which enables one
to analyze the equilibrium formation and filling of rings with a finite number of particles
interacting by means of Coulomb forces in the disk geometry.
Our approach is based on the cyclic symmetry and periodicity
of the Coulomb energy between particles located over different rings. As a result,
the problem of $n$ interacting charged particles is reduced to
the description of $p\,(\ll\!\! n)$ rings,
with homogeneously distributed integer charges.
 This picture is good enough
to obtain exact ground state configurations with correct energies,
excluding a few particular cases, up to $n\leq\!105$. 
For bigger systems the solution of the model equations provide also  
very good approximations to the exact ground state configurations. 
Indeed, the energy errors do not exceed a small percentage fraction of the exact values.
For $n\gtrsim 200$ our approach predicts the formation of 
the {\it hexagonal core} and {\it valence} circular rings for the centered configurations. 
The general evolution of the shell structure with an increasing 
number of particles is also properly described including finite size fluctuations. 
Note that the basic principles discussed for the disk geometry can also be applied to
parabolic confinement, or any other circular potential. 
The computational effort to get global energy minima is much less than in MD or simulated 
annealing calculations. In fact, simulation times in these methods can be 
drastically reduced by feeding them with initial configurations obtained 
by means of our method. 
Finally, we have analyzed and quantified the range of applicability of a pure 
classical picture for the description of charged particles under hard circular confinement.

\section*{Acknowledgments}
This work was supported in
part by Bogoliubov-Infeld program of BLTP and
RFBR (Russian Federation), Grant 14-02-00723.

\end{document}